\documentclass[manuscript,screen]{acmart}

\usepackage{multirow,tabularx,ragged2e}
\usepackage{subcaption}
\usepackage{todonotes}

\usepackage{enumitem}

\AtBeginDocument{%
  }

\setcopyright{acmcopyright}
\copyrightyear{2018}
\acmYear{2018}
\acmDOI{XXXXXXX.XXXXXXX}

\acmJournal{TOG}
\acmVolume{37}
\acmNumber{4}
\acmArticle{111}
\acmMonth{8}

\usepackage{color,soul}
\usepackage{framed}
\usepackage{ifthen}
\definecolor{shadecolor}{RGB}{180,180,180}

\newboolean{commentsenabled}
\setboolean{commentsenabled}{true}

\ifthenelse{\boolean{commentsenabled}}{
\newcommand{\unknown}[1]{\textcolor{blue}{Unknown says: {#1}}}
\newcommand{\anita}[1]{\textcolor{purple}{Anita says: {#1}}}
\newcommand{\martin}[1]{\textcolor{blue}{Martin says: {#1}}}
\newcommand{\gennady}[1]{\textcolor{blue}{Gennady says: {#1}}}
\newcommand{\kyoungsook}[1]{\textcolor{blue}{Kyoung-Sook says: {#1}}}
\newcommand{\john}[1]{\textcolor{blue}{John says: {#1}}}
\newcommand{\raymond}[1]{\textcolor{blue}{Raymond says: {#1}}}
\newcommand{\taylor}[1]{\textcolor{blue}{Taylor says: {#1}}}
\newcommand{\cyrus}[1]{\textcolor{blue}{Cyrus says: {#1}}}
\newcommand{\li}[1]{\textcolor{blue}{Li says: {#1}}}
\newcommand{\xiqi}[1]{\textcolor{blue}{Xiqi says: {#1}}}
\newcommand{\yang}[1]{\textcolor{blue}{Yang says: {#1}}}
\newcommand{\dimitris}[1]{\textcolor{blue}{Dimitris (Sacharidis) says: {#1}}}
\newcommand{\gabriel}[1]{\textcolor{blue}{Gabriel says: {#1}}}
\newcommand{\johannes}[1]{\textcolor{blue}{Johannes says: {#1}}}
\newcommand{\peer}[1]{\textcolor{blue}{Peer says: {#1}}}
\newcommand{\amr}[1]{\textcolor{blue}{Amr says: {#1}}}
\newcommand{\mahmoud}[1]{\textcolor{blue}{Mahmoud says: {#1}}}
\newcommand{\kristian}[1]{\textcolor{blue}{Kristian says: {#1}}}
\newcommand{\marc}[1]{\textcolor{blue}{Marc says: {#1}}}
\newcommand{\esteban}[1]{\textcolor{blue}{Esteban says: {#1}}}
\newcommand{\natalia}[1]{\textcolor{blue}{Natalia says: {#1}}}
\newcommand{\walid}[1]{\textcolor{blue}{Walid says: {#1}}}
\newcommand{\mario}[1]{\textcolor{brown}{Mario says: {#1}}}
\newcommand{\xu}[1]{\textcolor{blue}{Xu says: {#1}}}
\newcommand{\siva}[1]{\textcolor{blue}{Siva says: {#1}}}
\newcommand{\sarwat}[1]{\textcolor{blue}{Mo says: {#1}}}
\newcommand{\maxime}[1]{\textcolor{blue}{Maxime says: {#1}}}
\newcommand{\egemen}[1]{\textcolor{blue}{Egemen says: {#1}}}
\newcommand{\joonseok}[1]{\textcolor{blue}{Joon-Seok says: {#1}}}
\newcommand{\yannis}[1]{\textcolor{blue}{Yannis says: {#1}}}
\newcommand{\jianqiu}[1]{\textcolor{blue}{Jianqiu says: {#1}}}
\newcommand{\panos}[1]{\textcolor{blue}{Panos says: {#1}}}
\newcommand{\christian}[1]{\textcolor{blue}{Christian says: {#1}}}
\newcommand{\mengxuan}[1]{\textcolor{blue}{Mengxuan says: {#1}}}
\newcommand{\flora}[1]{\textcolor{magenta}{Flora says: {#1}}}
\newcommand{\bettina}[1]{\textcolor{blue}{Bettina says: {#1}}}
\newcommand{\jussara}[1]{\textcolor{blue}{Jussara says: {#1}}}
\newcommand{\sanjay}[1]{\textcolor{blue}{Sanjay says: {#1}}}
\newcommand{\reynold}[1]{\textcolor{blue}{Reynold says: {#1}}}
\newcommand{\dimitrios}[1]{\textcolor{blue}{Gunopulos says: {#1}}}
\newcommand{\mokbel}[1]{\textcolor{blue}{Mokbel says: {#1}}}
\newcommand{\matthias}[1]{\textcolor{teal}{Matthias says: {#1}}}
\newcommand{\carola}[1]{\textcolor{blue}{Carola says: {#1}}}
\newcommand{\moustafa}[1]{\textcolor{blue}{Moustafa says: {#1}}}
\newcommand{\demetris}[1]{\textcolor{blue}{Demetris says: {#1}}}
\newcommand{\goce}[1]{\textcolor{violet}{Goce says: {#1}}}
\newcommand{\song}[1]{\textcolor{blue}{Song says: {#1}}}
\newcommand{\andi}[1]{\textcolor{red}{Andreas says: {#1}}}
}
{
\newcommand{\unknown}[1]{}
\newcommand{\anita}[1]{}
\newcommand{\martin}[1]{}
\newcommand{\gennady}[1]{}
\newcommand{\kyoungsook}[1]{}
\newcommand{\john}[1]{}
\newcommand{\raymond}[1]{}
\newcommand{\taylor}[1]{}
\newcommand{\cyrus}[1]{}
\newcommand{\li}[1]{}
\newcommand{\xiqi}[1]{}
\newcommand{\yang}[1]{}
\newcommand{\dimitris}[1]{}
\newcommand{\gabriel}[1]{}
\newcommand{\johannes}[1]{}
\newcommand{\peer}[1]{}
\newcommand{\amr}[1]{}
\newcommand{\mahmoud}[1]{}
\newcommand{\kristian}[1]{}
\newcommand{\marc}[1]{}
\newcommand{\esteban}[1]{}
\newcommand{\natalia}[1]{}
\newcommand{\walid}[1]{}
\newcommand{\mario}[1]{}
\newcommand{\xu}[1]{}
\newcommand{\siva}[1]{}
\newcommand{\sarwat}[1]{}
\newcommand{\maxime}[1]{}
\newcommand{\egemen}[1]{}
\newcommand{\joonseok}[1]{}
\newcommand{\yannis}[1]{}
\newcommand{\jianqiu}[1]{}
\newcommand{\panos}[1]{}
\newcommand{\christian}[1]{}
\newcommand{\mengxuan}[1]{}
\newcommand{\flora}[1]{}
\newcommand{\bettina}[1]{}
\newcommand{\jussara}[1]{}
\newcommand{\sanjay}[1]{}
\newcommand{\reynold}[1]{}
\newcommand{\dimitrios}[1]{}
\newcommand{\mokbel}[1]{}
\newcommand{\matthias}[1]{}
\newcommand{\carola}[1]{}
\newcommand{\moustafa}[1]{}
\newcommand{\demetris}[1]{}
\newcommand{\goce}[1]{}
\newcommand{\song}[1]{}
\newcommand{\andi}[1]{}
}

\begin{document}

\title{Mobility Data Science: Perspectives and Challenges}
\titlenote{This is a community publication. The authors of this article met in Dagstuhl for Seminar \#22021 on Mobility Data Science~\cite{mokbel2022mobility}. The first four authors co-organized the Dagstuhl seminar leading to this article and coordinated the creation of this manuscript. All other authors contributed equally to this research. The seminar was held in the week of January 9 - 14 , 2022. It had 47 participants specialized in different topics: data management, mobility analysis, geography, privacy, urban computing, systems, simulation, indoors, visualization, information integration, and theory. Due to COVID-19, the seminar took place in hybrid mode, with 8 onsite, and 39 remote participants. Despite the challenge of different time zones of the participants, all sessions were attended by at least 37 participants. It was an excellent opportunity to start the discussion about next-decade opportunities and challenges for mobility data science.}

\author[1]{Mohamed F. Mokbel}
\author[2]{Mahmoud Sakr}
\author[3]{Li Xiong}
\author[4]{Andreas Züfle}
\author[5]{Jussara Almeida}
\author[4]{Taylor Anderson}
\author[6]{Walid G. Aref}
\author[7]{Gennady Andrienko}
\author[7]{Natalia Andrienko}
\author[8]{Yang Cao}
\author[9]{Sanjay Chawla}
\author[10]{Reynold Cheng}
\author[11]{Panos K. Chrysanthis}
\author[4]{Xiqi Fei}
\author[12]{Gabriel Ghinita}
\author[13]{Anita Graser}
\author[14]{Dimitrios Gunopulos}
\author[15]{Christian S.~Jensen}
\author[16]{Joon-Seok Kim}
\author[17]{Kyoung-Sook Kim}
\author[18]{Peer Kr\"oger}
\author[19]{John Krumm}
\author[20]{Johannes Lauer}
\author[21]{Amr Magdy}
\author[22]{Mario A. Nascimento}
\author[23]{Siva Ravada}
\author[18]{Matthias Renz}
\author[2]{Dimitris Sacharidis}
\author[25]{Flora Salim}
\author[26]{Mohamed Sarwat}
\author[2]{Maxime Schoemans}
\author[24]{Cyrus Shahabi}
\author[27]{Bettina Speckmann}
\author[28]{Egemen Tanin}
\author[29]{Xu Teng}
\author[30]{Yannis Theodoridis}
\author[15]{Kristian Torp}
\author[29]{Goce Trajcevski}
\author[31]{Marc van Kreveld}
\author[32]{Carola Wenk}
\author[33]{Martin Werner}
\author[34]{Raymond Wong}
\author[2]{Song Wu}
\author[35]{Jianqiu Xu}
\author[36]{Moustafa Youssef}
\author[37]{Demetris Zeinalipour}
\author[30]{Mengxuan Zhang}
\author[2]{Esteban Zim\'anyi}

\renewcommand{\shortauthors}{Mokbel, Sakr, Xiong, Z\"ufle et al.}

\makeatletter
\let\@authorsaddresses\@empty
\makeatother

\begin{abstract}
  Mobility data captures the locations of moving objects such as humans, animals,  and cars. With the availability of GPS-equipped mobile devices and other inexpensive location-tracking technologies, mobility data is collected ubiquitously. In recent years, the use of mobility data has demonstrated significant impact in various domains including traffic management, urban planning, and health sciences. In this paper, we present the domain of mobility data science. Towards a unified approach to mobility data science, we present a pipeline having the following components: mobility data collection, cleaning, analysis, management, and privacy. For each of these components, we explain how mobility data science differs from general data science, we survey the current state of the art, and describe open challenges for the research community in the coming years. 
\end{abstract}

\maketitle

\sethlcolor{shadecolor}

    \begin{figure*}[t]
    \centering
    \includegraphics[width=0.99\columnwidth] {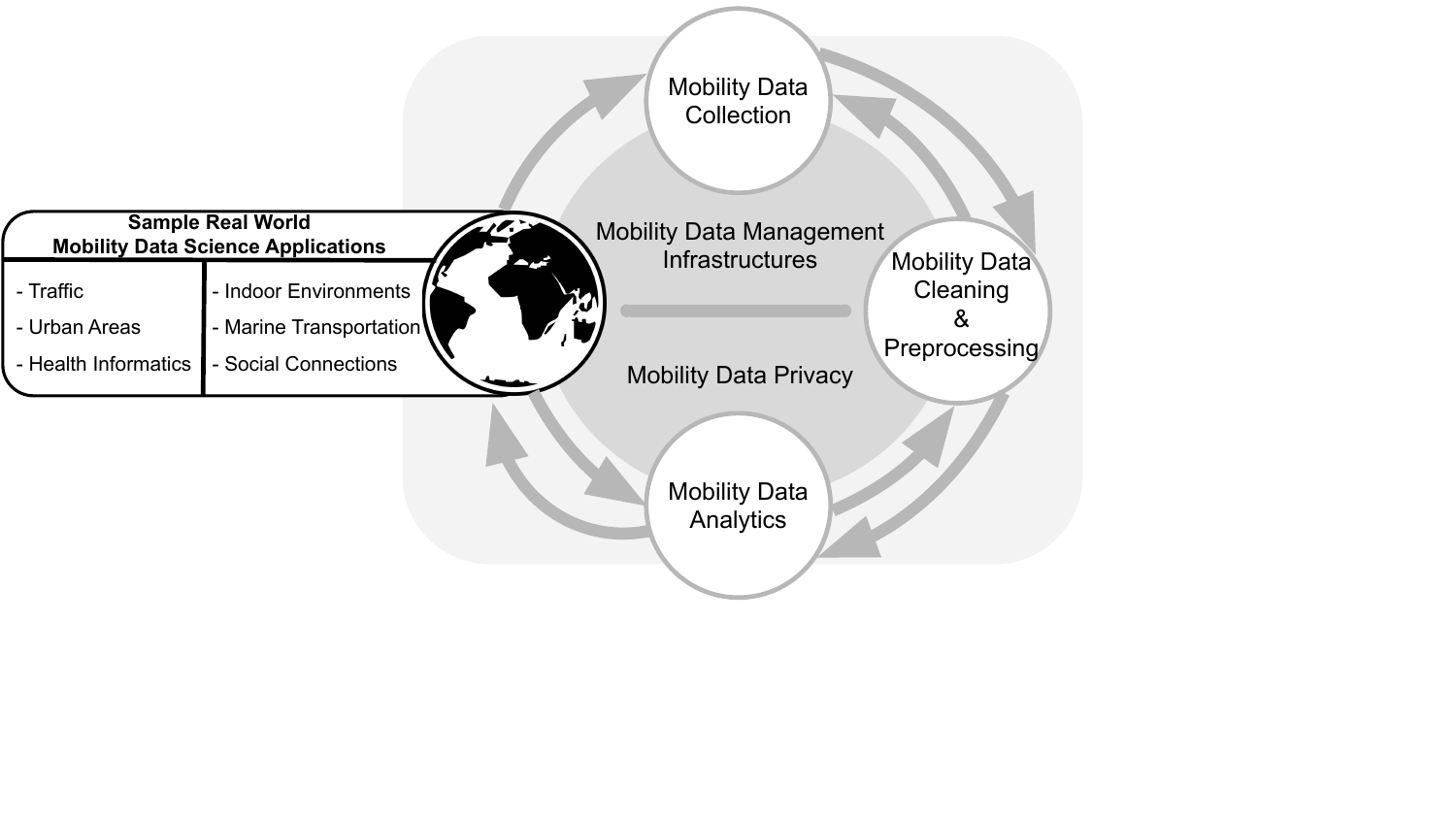}    \vspace{-0.5cm}
    \caption{The Mobility Data Science Pipeline \vspace{-0.5cm}}
    \label{fig:dspipeline}
    \end{figure*}

\section{Introduction}


The volume of mobility data being collected has been steadily increasing since the advent of affordable personal location-enabled mobile devices. Examples of mobility data continuously generated and collected in huge volumes include: (a)~individual sporadic locations obtained from mobile app data and location-based social networks, (b)~individual pedestrians, biking, or driving trajectories constrained by underlying side walks, biking trails, and road network, respectively, 
(c)~indoor individual or asset tracking data obtained from RFID and bluetooth devices, (d)~athletes movement data in various sports obtained from wearable devices, (e)~public transportation, taxis, ride sharing, and delivery logistics trajectories obtained by location-tracking devices and specially designed app services, (f)~aircraft and vessels trajectories moving in an unconstrained environment (i.e., no underlying road network) obtained by air and sea traffic monitoring services, and (g)~animal tracking data moving freely in the space obtained from physically tagged and remotely sensed animals. Generally speaking, for each moving object, mobility data is typically available in the form of a sequence of ({\em location}, {\em timestamp}) pairs. The {\em location} attribute could be as simple as a point, represented by either {\em latitude} and {\em longitude} coordinates or as relative coordinates with respect to the underlying space. The {\em location} attribute could also be an area, which can represent the mobility of objects with spatial extents, e.g., flocks or group movement.


The ability of understanding and analyzing mobility data is crucial for various widely used important sectors and applications. In transportation and traffic management, analyzing traffic data through vehicle mobility helps in predicting accidents~\cite{moosavi2019accident}, traffic congestions~\cite{yuan2021survey}, and better route planning~\cite{chen2019effective}. In ride sharing and delivery logistics application, analyzing trip mobility data help in data-driven eco route planning, which results in huge cost and energy savings~\cite{guo2015ecomark}. In location-based services, analyzing people movements around the city significantly helps in trip planning activities~\cite{teng2020semantically}, finding popular tourists sites and restaurants~\cite{josse2015tourismo}, and data-driven routing and querying~\cite{teng2021semantically}. In indoor navigation, understanding how people move indoors helps in understanding the traffic for various stores inside a mall, which is needed in various market research studies~\cite{jensen2010indoor}. In urban planning, driving data can significantly help in building highly accurate, reliable, and annotated maps~\cite{musleh2021qarta} as well as deciding on good location for various facilities, e.g., restaurants, retail stores, and clinics~\cite{shang2011finding}. In social computing, analyzing how people move in cities and regions helps in understanding the demand for infrastructure and energy as a means of reducing inequalities~\cite{salimBAE}. In disaster response, analyzing crowd movement helps in preparing for natural disasters through rescuing and evacuation efforts~\cite{hong2021measuring}. In health informatics, connected wearables can monitor and analyze the movement of elderly people, allowing for timely, and potentially life-saving, interventions~\cite{lin2020healthwalks}. In pandemic prevention, the ability of privacy-preserving individual tracking allows for contact tracing, which was deemed as cornerstone in limiting pandemic spread ~\cite{mokbel2022introduction,zufle2022introduction}.


Despite the common goal of acquiring, managing, and generating insights from mobility data, the mobility data science community is largely fragmented, developing solutions in silos. It stems from a range of disciplines with expertise in moving object data storage and management \cite{guting2005moving}, geographic information science \cite{GIScience10}, spatiotemporal data mining \cite{ijgi4042306}, human mobility modelling \cite{BARBOSA20181}, ubiquitous computing, computational geometry and more.
The sheer volumes of mobility data along with the immense need of mobility data analysis in various applications call for employing a complete Data Science pipeline~\cite{Rezig19} over mobility data. This includes the whole pipeline of Data Science applications, starting from the data storage and management infrastructure and going through data collection, data cleaning and preprocessing, and data analysis. Unfortunately, this is not straightforward as current Data Science systems, tools, and algorithms are not directly applicable to mobility data. This is mainly due to the fact that these systems, tools, and algorithms, are designed in a generic way to support any data type, and hence they do not lend themselves to the distinguishing characteristics of mobility data. Examples of such characteristics include the spatial and temporal dimensions of the data, the rate of updates, and the privacy requirements. In particular, mobility data is always spatial, where nearby objects are more related to each other. This is unlike traditional data, where the concept of nearby and locality is not taken into account. Also, similar to time series data, mobility data is temporal, where one object may have hundreds of updates to its location, and all updates are related to each other (e.g., one trajectory). This is again unlike traditional data, where temporal updates of a single object are not frequent and older updates would be of less importance. Similar to streaming data, mobility data has a high frequency of updates, which is not supported in typical data science applications. Finally, mobility data is more sensitive to privacy. While privacy-preserving in traditional data can be achieved by removing (quasi-)identifier attributes, in mobility data, locations by themselves are considered private information that can reveal not only the users' identities, but also their behavior, life style, medical conditions, and work places.


Motivated by ubiquity and sheer volumes of mobility data, the importance of mobility applications, and the lack of support from current data science pipelines, this paper presents a pipeline for Mobility Data Science. We define \emph{Mobility Data Science} as an interdisciplinary field that uses scientific methods, processes, algorithms and systems to extract or extrapolate knowledge and insights from potentially noisy, structured and unstructured mobility data, and apply knowledge from mobility data across a broad range of application domains. While currently, the community of developers, practitioners, and researchers, dealing with mobility data use off-the-shelf data science techniques and systems to collect, clean, manage, and analyze their mobility data, we firmly believe that this ends up to sub-bar performance. We urge such community to build its own mobility data science pipeline to better serve its own purpose. This paper makes the case for the need for a mobility data science pipeline along with the challenges that need to be addressed to realize it.


\section{Mobility Data Collection}
\label{subsec:acqusition}

The abundance availability of real data is a cornerstone to any data science application, and mobility data science applications are of no exception. However, it is much easier to collect and find tons of data for data science applications than it is the case for mobility data science. In particular, for data science applications, well-established research in anonymizing personal data allows wide data sharing. This is to the extent that governments have released various datasets for public (e.g., Data.gov). In addition, companies already collect their own inventory data that does not include any personal identifiers, and hence it is suitable to be fed to data science applications. On the other side, data-driven mobility data science research has been in a constant struggle with the need for available mobility data. A main reason is that non-aggregated individual human location data is considered personal identifiable information as it may lead to tracing an individual's identity. For example, it has been shown that only a few spatial locations are sufficient to uniquely identify individuals even among a large population of people~\cite{seglem2017privacy}. As a result, most datasets are collected in aggregated form, which hinders the deployment of various mobility data science applications. This sections discuss current efforts and challenges of mobility data collection.

\subsection{Efforts in Mobility Data Collection}

Before the wide availability of personal digital devices, human mobility data collection was expensive and therefore datasets were very sparse. With the advent of personal location-enabled devices, many people's movements have started leaving digital traces that are being collected either by industry as a means of providing location-based services~\cite{safegraph} or by governmental entries as a means of data analysis, e.g., traffic-related studies~\cite{usdotdata}. However, this did not result in a similar explosion of publicly available mobility data, mainly due to privacy and data sharing concerns.

Current efforts in releasing public non-aggregated mobility data is mainly limited to small datasets, small regions, while removing locations that can lead to one's whereabouts. This mostly include trips obtained from taxis, ride sharing services, or public transportation. Some of these datasets include detailed trajectory data for the following cities (ordered alphabetically): (1)~{\em Athens}~\cite{BG20}. 500K trajectories collected over 5 days in downtown Athens, Greece, (2)~{\em Beijing~1}~\cite{ZXM10}. 17+K trajectories with 26 Million GPS points over three years in Beijing, China, (3)~{\em Beijing~2}~\cite{yuan2010t}. 10+K trajectories with 15 Million GPS points over one week in Beijing, China, (4)~{\em Rio}~\cite{DC18}. 12+K Buses with detailed trajectories of 118+ Million GPS points over 30 days in Rio de Janeiro, Brazil, (5)~{\em Rome}~\cite{BBL+14}. 320 Taxis with detailed trajectories of 21+ Million GPS points over 30 days in Rome, Italy, (6)~{\em San Francisco}~1~\cite{PSG09}. 536 Taxis with detailed trajectories of 11+ Million GPS points over 30 days in San Francisco, CA, USA, (7)~{\em San Francisco}~2~\cite{GISCup17}. 20+K detailed trajectories with 5+ Million GPS points in San Francisco, CA, USA. (8)~{\em Shenzhen}~\cite{WCZ+19}. 664 Taxis with detailed trajectories of 1.1+ Millions GPS points over one day in Shenzhen, China, (9)~{\em Singapore}~\cite{HYL+19}. 84K trajectories with 80+ Million GPS points over a month in Singapore. Other datasets only include the origin and destination of each trajectory. Examples include the following cities: (1)~{\em Austin}~\cite{Austin}. 1.5 Million trips for a period of 10 months in Austin, TX, USA, (2)~{\em Guangdong}~\cite{YZZ18}. 2.5 Million trips over one day in Guangdong Province, China, (3)~{\em NYC}~\cite{NYC}. 1.5 Million Taxi trips over a period of 6 months in New York City, NY, USA, (4)~{\em Porto}~\cite{Porto}. 426K Taxi trips over three months in Porto, Portugal,

Other than trip and trajectory road network data, there are tons of available biking data across the world, including tens of millions trips in Bay Area~\cite{BayAreaBikes}, Boston~\cite{BostonBikes}, Chicago~\cite{ChicagoBikes}, Columbus~\cite{ColumbusBikes}. London~\cite{LondonBikes}, Los Angeles~\cite{LosAngelesBikes}, Madrid~\cite{MadridBikes}, Minneapolis~\cite{MinneapolisBikes}, New York City~\cite{NYCBikes}, Philadelphia~\cite{PhiladelphiaBikes}, Portland~\cite{PortlandBikes}, and Washington D.C.~\cite{WashingtonBikes}. There are also available public marine traffic that include detailed vessel trajectories (e.g.~\cite{patroumpas2017online}), sport data sets for basketball and soccer that include a variety of events took place in major leagues within one season~\cite{pappalardo2019public}, and indoor data about the behaviors of nearly 30 Year-10 students and their teachers collected over four weeks in Australia, with spatial reference (associations to rooms) and highly granular wearable data~\cite{gao2022understanding}.

However, there are some large-scale aggregated datasets on a coarse granularity that can help in high level analysis, but not to get insight details of mobility data. Examples of such aggregate data include origin-destination employment statistics in the USA that contains home-to-work commuting flows aggregated to census tract level~\cite{graham2014design}, cell phone trace datasets capturing the locations of individuals aggregated to their nearest cell tower~\cite{vigfusson2021cell}, foot traffic data of check-ins of 35 million anonymized mobile devices in USA are aggregated to census block groups~\cite{safegraphweeklypatterns}, and a global database about aggregate indoor occupant behavior, composed of 34 datasets from 15 countries and 39 institutions, collected by occupancy sensors that measure the occupancy count of each space being monitored~\cite{dong2022global}. An additional source of human mobility data is location-based social network (LBSN) data. LBSN data captures both 1) discrete check-ins between users and locations, and 2) a social network between users. This dimension of location bridges the gap between the physical world and online social networking services~\cite{zheng2011location}. However, it has been shown existing LBSN data sets are too small to broadly understand, analyze, and predict human behavior~\cite{li2016assessing}.

The lack of available mobility data, combined with the need to stress test various research ideas have motivated various research groups to either develop their own data simulators or develop publicly available simulators that can also be used by other researchers for benchmark datasets. However, such simulators were mainly designed to test specific aspects of research, but not meant to be representative of real mobility data. For example, various simulators were mainly designed to test new index structures for mobility data, query processing algorithms, and system infrastructure scalability for managing spatiotemporal data (e.g.,~\cite{mokbel2013mntg}). Within the transportation community, more fine granularity simulators (e.g.,~\cite{sumo}) were proposed to study traffic infrastructure, but none of them is meant to provide comprehensive mobility study.

\subsection{Challenges in Mobility Data Collection}

This section presents some of the challenges in mobility data collection that the community needs to address towards realizing the pipeline of mobility data science.

\paragraph{Challenge 1. Mobility Data Privacy.} In most cases, (human) mobility data is sensitive and considered as personal identifiable information. This puts major privacy concerns on data sharing. Hence, any attempt to collect fine granularity detailed trajectory or human mobility data must first address the privacy challenge. Though the general topic of data privacy has been well studied in literature with practical solutions, such solutions are not directly applicable to the case of mobility data. In particular, mobility data gives rise to the TUL (Trajectory-User Linking) problem~\cite{gao2017identifying}. To protect users' actual locations, while preserving meaningful mobility information for various learning tasks, one may wish to generate realistic motions based on real-world mobility datasets~\cite{ZhouLZT2021-Traces}. Since privacy is a core problem in mobility data that does not only impact data collection, but also impacts all other components of the mobility data science pipeline, we dedicate Section~\ref{sec:privacy} to discuss mobility data privacy in details.

\paragraph{Challenge 2. Mobility Data Bias.} Mobility data collection procedures suffer from all kinds of bias. For example, mobile applications data and mobile phone network data are biased against people who do not use smart phones or use prepaid plans. Most traffic counting sensors are installed to count cars but do not count pedestrians, cyclists, wheelchairs, or similar. Cells in mobile phone networks vary widely in size. The data traces that are usually collected in cellular networks are cellular themselves. This affects rural areas with larger cells more than urban areas. Volunteered tracking data is biased towards technically savvy people. Sports tracking data is biased towards health conscious middle and upper class. It is important to understand, measure, and mitigate data bias in mobility datasets to ensure that actions and policies that are based on mobility data science results are equitable, fair, and include vulnerable populations~\cite{ShahamGS22}.

\paragraph{Challenge 3. Incentives for Data Sharing.} Users need to have good incentives to share their locations. To some degree, users kind of agree to share their locations with commercial entities to get location-based services, ride sharing, cell phone coverage, delivery, among other services. However, it is understood that users would be reluctant to publicly share their mobility traces. Meanwhile, the biking community have shown a great example for sharing their biking trails. A main reason is that, in many places of the world, most of these trails are not really home-to-work commuting, but it is more of an outdoor activities. Hence, sharing biking trails helps fellow bikers in knowing the conditions of biking trails, which is a great incentive for sharing. More incentives need to be given for drivers to share their mobility traces, even for sporadic trips that do not lead to identifiable locations. Sharing could be for part of the trajectory where rewards are given back based on the sharing length and resolution. A gamification concept may be exploited to encourage more participants to share.

\paragraph{Challenge 4. Simulated Mobility Data.} The dire need to mobility data along with the difficulty of obtaining them made it apparent that simulated synthetic data is immensely needed to enrich and train mobility data science applications. However, the challenge is to go beyond earlier attempts of simulating data for testing very specific techniques to simulating data for the general purpose of having realistic life scenarios. Empowered by modern computational capabilities that make it possible to simulate large populations, the mobility community should work with social scientists to create realistic individual-level human mobility data. Lessons have been learned from the experience of the deep learning community, by applying generative adversarial networks (GANs) for trajectory generation~\cite{zhang2022factorized}. However, it is unclear as of yet, how to measure the realism of mobility data. If synthetic mobility data is too realistic, for example, due to training on real human trajectories, it may invade someone's privacy if, for instance, it shows where members of a given household actually visit. On the flipside, benchmark data that is too disconnected from the real-world and does not represent realistic human behavior would not allow to generalize to the real-world.

\clearpage
\section{Mobility Data Cleaning}
\label{sec:cleaning}

Until the early 21st century, location data and mobility data available for geographic information science (GIS) was mainly
collected, curated, standardized \cite{fegeas1992overview,FGDC}, and published by authoritative sources
such as the United States Geological Survey (USGS)~\cite{USGS}. Now, data used for mobility data science is often obtained from sources of volunteered geographic information (VGI)~\cite{sui2012crowdsourcing}. Such data is contributed by millions of individual users (more than ten million contributors in the case of OpenStreetMap~\cite{OSM}) and is rarely curated. Mobility data collected from such sources is highly uncertain due to physical limitations of sensing devices, due to obsoleteness of observations, and in many cases plain incorrect due to deliberate misinformation~\cite{mooney2017review}.
Consequentially, our ability to unearth valuable knowledge from large sets of mobility data is often impaired by the uncertainty of the data which geography has been named the ``the Achilles heel of GIS'' \cite{goodchild1998uncertainty}.

Data cleaning and preprocessing is a milestone to all data science. In fact, it has been reported that data scientists spend more than 80\% of their time in data cleaning and preparations~\cite{data_cleaning_nyt}. As a result, there are huge efforts in the data science community dedicated to developing various data cleaning algorithms~\cite{chu2016data} and full-fledged systems~\cite{dallachiesa2013nadeef}. Mobility data is of no exception in terms of its need for data cleaning and preparation procedures. But for numerous reasons, data cleaning and preparation yields unique challenges.
This section discusses current efforts and challenges of mobility data cleaning.

\subsection{Efforts in Mobility Data Cleaning}
A recent survey~\cite{LLJ+23} and data quality assessment tool~\cite{Graser21} have discussed various sorts of errors that negatively impact data quality in spatial and mobile environments.
Motivated by the inaccuracy of location tracking devices, several efforts were dedicated to address: (a)~the spatial inherent inaccuracy of GPS devices and (b)~the uncertainty of moving objects whereabouts between each two known locations, which is a result of low sampling rates due to bandwidth and battery limitations.

As the spatial inaccuracy indicates erroneous GPS coordinates, the efforts to identify and correct such coordinates have focused on either finding and eliminating outliers or map matching all coordinates to an underlying fixed and trusted infrastructure (e.g., road network map). For the case of map matching, existing efforts aim to match/snap all GPS traces to an underlying road network~\cite{BPS+05,CXH+20}. Proposed techniques vary from as simple as snapping each point to its nearest road to applying Markov Chain to identify the most probable road segment that each point should be snapped to. In case there is no underlying road infrastructure (e.g., marine transportation or animal movement), outlier detection techniques are used to identify and remove erroneous points~\cite{toohey2015trajectory}.

Irrespective of the collection method and device settings, there is also indispensable uncertainty in movement data caused by their discreteness.  Since time is continuous, the data cannot refer to every possible instant. For any two successive instants, there is a temporal gap where the whereabouts of the moving objects are unknown. To overcome such location uncertainty, several efforts were dedicated to modeling the uncertainty of mobility data surveyed in~\cite{zufle2017handling}.

\clearpage
\subsection{Challenges in Mobility Data Cleaning}

This section delves into some challenges linked to cleaning mobility data that the community need to tackle.

\vspace{2pt}
\paragraph{Challenge 5. Inaccuracy in the Movement Space Infrastructure.} A unique challenge in mobility data is that in many cases, its reference points are the ones that are inaccurate. In particular, mobility data that represent movement on a road network may be more accurate than the road network itself. Road networks, like any other type of data, suffer from all sorts of inaccuracy, and may not be even available in many places~\cite{MM22A}. In fact, Microsoft has recently announced that it has found more than one million kilometers of roads missing from current maps~\cite{MicrosoftMissingRoads}. This is why there is a whole area of industrial and academic research about map inference, which aims to infer (all or missing parts) of the road network from either satellite images~\cite{BHA+18} or trajectory data~\cite{BE12}. However, almost all of these techniques focus on making accurate maps in terms of topology. There need to be more efforts on map inference algorithms that go beyond inferring the map topology to inferring map metadata (e.g., road speed, traffic lights, number of lanes, and turns), without which, mobility data would not be accurate as its road network reference itself is missing important data. A major step towards cleaning mobility data would be to first clean its reference map.

\paragraph{Challenge 6. Filling in Temporal Mobility Gaps.} As mentioned earlier, there are lots of efforts dedicated to modeling the uncertainty of moving objects whereabouts between each two consecutive time instances. However, uncertainty poses different challenges to down stream functions and applications, including the need to develop new techniques for indexing, query processing, and data analysis for various uncertainty models. One way to overcome this is to try to infer the actual whereabouts of a moving objects between any two time instances with known locations. There are already several efforts to insert artificial points between each two consecutive trajectory points, with the promise that these points act as if the trajectory was collected in a very high sampling rate. This process has various names, e.g., trajectory interpolation~\cite{Lon16,ZZX+12}, trajectory completion~\cite{LLG+16}, trajectory data cleaning~\cite{ZSW+17}, trajectory restoration~\cite{LCK+21}, trajectory map matching~\cite{BPS+05}, trajectory recovery~\cite{WWL+21}, and trajectory imputation~\cite{EIM22}. However, the large majority of such work rely on matching the trajectory points on the underlying road network, where the imputation becomes finding the road network shortest path between each two consecutive trajectory points. Unfortunately, this is not applicable to the case where the road network is unknown, untrusted, or inaccurate. Hence, more recent attempts try to do data-driven trajectory imputation without relying on the underlying road network~\cite{EIM22,FLZ+18}. However, these techniques are either not scalable to city-scale trajectory datasets,  or require 
dense historical data that derives its imputation process. There is an immense need to develop a scalable, accurate, and fine-grained imputation that almost mimics a continuous data stream of trajectory locations.

\clearpage
\section{Mobility Data Analytics}\label{sec:analysis}

Spatial data is special. Unlike non-spatial features, location attributes (e.g.,  longitude and latitude) rarely exhibit linear or other simple functional relationships to variables of interest. It rarely makes sense to model a variable of interest directly in relation to spatial attributes. Instead, it is distances that matter. According to Tobler’s first rule of Geography, ``everything is related to everything else, but closer things are more related than things that are far apart''~\cite{tobler1970computer}. For mobility data, proximity is further extended with time, i.e., objects that are close in space and time. In addition to this concept of spatiotemporal autocorrelation, what makes mobility data even more challenging to handle is that it is often observed from humans whose behavior can often be irrational and difficult to explain. As Nobel Prize laureate Murray Gell-Mann famously said, ``Think how hard physics would be if particles could think''~\cite{page1999computational}. But unlike in physics, the ``particles'' of interest are often humans who can think. Data collection sensors have the capability to capture the spatiotemporal locations of the moving objects, but not their behavioral aspects. These difficulties require new paradigms, techniques, and algorithms to analyze and learn from the spatiotemporal data, and that can explain and predict the associated behavior.  This section discusses current efforts and challenges of mobility data analysis.


\subsection{Efforts in Mobility Data Analytics} 

Mobility data analytics has already gained momentum in research in the recent years. Dedicated workshops have existed in major conferences; including the ACM SIGSPATIAL International Workshop on Analytics for Big Geospatial Data (BigSpatial) since 2011~\cite{shashidharan20219th}, the Big Mobility Data Analytics (BMDA) workshop in EDBT since 2018~\cite{DBLP:journals/geoinformatica/X22}, and the ACM SIGSPATIAL International Workshop on Animal Movement Ecology and Human Mobility 
(HANIMOB)@SIGSPATIAL since 2021~\cite{ossi2022hanimob}. Surveys on the status of research exist~\cite{10.1145/3161602, 10.1007/s10707-022-00483-0}.  

Mobility data analytics encompasses various application domains and involves analyzing data from different -- sources such as urban~\cite{zhao2016urban}, maritime~\cite{claramunt2017maritime}, aviation~\cite{chung2020data}, animal movement~\cite{ossi2022hanimob}, and indoor movement~\cite{jensen2010indoor}. Among these different themes urban mobility stands out with a fairly large body of research including green routing~\cite{almeida2017gps2gr}, traffic anomaly detection~\cite{pan2013crowd}, hot spot and hot path  analysis~\cite{Panagiotis18}, road traffic prediction~\cite{NAGY2018148}, and travel time estimation~\cite{wang2019simple}. Trajectories of moving objects have been used as means to create and continuously update the road network~\cite{musleh2021qarta}. Public transport systems also collect ticketing data in the form of passenger check-ins, sometimes also associated with check-outs. This data has been shown very useful to transit planners in understanding passenger demand and movement patterns in daily operations as well as in the strategic long-term planning of the network~\cite{truong2018towards}. Personal mobility of individuals is also a subject of analysis that includes analyses, e.g.,  activity recognition~\cite{parent2013semantic,Chen22}, personalized routing \cite{Dai15}, matching with ride-sharing services~\cite{asghari2016price}, and crowd-sourcing~\cite{Pfoser2016}. 

While a significant portion of research focuses on understanding and analyzing data through analytics, there are also important efforts dedicated to developing generic analysis tools for spatiotemporal data that are agnostic to the application domain.
Efforts on generic methods for mobility data analysis include, among many others, trajectory clustering~\cite{Wang22}, trajectory similarity measures \cite{toohey2015trajectory}, outlier detection~\cite{han2022deeptea}, transportation mode classification~\cite{biljecki2013transportation}, spatiotemporal pattern detection \cite{sakr2014group}, and trajectory completion~\cite{krumm2022maximum}. However, and despite these many research efforts towards analyzing mobility data, there is lack of common data analysis tools and systems. The scientific software environment for mobility data analysis is rather fragmented. For example, \cite{joo_navigating_2020} lists 58 packages in their review of R packages for movement and \cite{graser_state_2023} reviews Python libraries for movement data analysis and visualization. 

Recent years have seen a notable increase on research on deep learning for mobility data analysis~\cite{LucaBLP23-Survey-Mobility,XieSLZYZMQZG18-Big-Mobile-Survey}. This brought an increased adoption of various paradigms and (adapted versions of) architectures used in other areas where deep learning has brought improvements in tasks, e.g., clustering/classification~\cite{MinGLZCL18-Clustering-Survey}, prediction~\cite{Lara-BenitezCR21-survey-time-series-deep-learning} and recommendation~\cite{BatmazYBK19-recommender-survey}, information propagation~\cite{ZhouXTZ21-cascades-survey}, etc. 
For example, Generative Adversarial Network (GAN) based architectures have been used recently to learn representations of trajectories and generate synthetic trajectories techniques~\cite{gao2022generative}. Given the introduction of Transformers~\cite{vaswani2017attention}, transformed-based approaches have also been used for mobility modelling and trajectory prediction~\cite{xue2021termcast}, given the sequential properties of mobility data. Other deep learning approaches such as contrastive learning~\cite{0002WXTT22-TIST} have also been exploited in mobile data settings, along with investigation of the impact/benefits of representation learning~\cite{GaoZZTYL22-INFSCI}.


\subsection{Challenges in Mobility Data Analysis} 

This section highlights open problems related to mobility data analysis, that needs consideration from the community.

\paragraph{Challenge 7. ML for Mobility Data.}
The state-of-the-art deep learning (DL) models, such as Transformers~\cite{vaswani2017attention} were not developed initially for mobility data science in mind. They were derived from NLP and computer vision domains. The community needs to provide best-case practices for doing ML (and DL) for mobility data. 

A major hurdle, and a research opportunity as well, is that existing ML and analytics tools, e.g., TensorFlow, or PyTorch, do not support location and mobility as base data types to reason about. So, even the basic analysis, such as clustering, classification, similarity, etc, need to be extended when mobility data is involved. These tasks, as well as higher-level analysis, can not be totally independent. Instead, common basic building blocks could have an impact on all or some of them. For example, exploring the effectiveness of embedding for mobility data analysis is a basic block that could impact different ML-based analysis tasks. This raises a challenge to build analysis primitives and common building blocks for applications that could shape a framework of ML-based mobility data analysis.

Another major hurdle is the robustness in data-driven mobility models.
It is widely known that data-driven models (as in the case of ML or DL) are only as good as the data that it is used to be trained on. However, given the changes of mobility behaviors, such as the COVID-19 pandemic and the associated lockdowns, and environmental events and disasters, traditional ML based, and even recent DL based, methods are no longer robust. The models' performance deteriorate in unseen events, especially as new behaviors emerge and then persist. Recent effort includes the incorporation of `contextual-awareness' and `memory' in an enhanced event-aware spatiotemporal network~\cite{wang2022event} for predicting mobility in multiple modes of transportation including taxi, cycling, subway during the unprecedented events like COVID lockdowns, or snowstorms, as it emerged and up to 30 days post the event. However, more work to be done on modelling and understanding mobility behavior, that are robust to changes due to societal events. 

\paragraph{Challenge 8. Progressing from Next Location Prediction to Movement Behavior Understanding.}
Due to the wide availability of aggregated check-in and foot-traffic data, many researchers focus on the problem of location prediction, e.g.,~\cite{xue2021mobtcast}. Leveraging predictions such as ``User X will visit Coffee Shop A next'' or ``$32\pm 4$ users will visit Coffee Shop A in the next hour'' has some direct applications. It could be useful for providing information about parking ``parking at location X appears to be a problem today, so consider ...'', for battery charging opportunities, or for providing information about collective transportation status ``Metro station X that you are expected to visit is closed for repairs, so instead ...''. One could provide a new transportation schedule and departure time in response to problems at an anticipated future location of a user, just like airlines at times update your itinerary in case of issues. Earlier work has been based on data mining techniques to detect periodic behavior, e.g.,~\cite{DBLP:conf/kdd/JindalGTLH13, DBLP:conf/edbt/ElfekyAE04, DBLP:conf/pkdd/BerberidisVAAE02}.
Beyond predicting locations, if we understand the underlying behavior, at the individual-, group-, or population-scales, that leads to these predictions, we could understand \textit{why} one coffee shop chain has
increasing visitor rates (e.g., due to a movement towards organic coffee sold by the coffee shop). 
Through inferring from the data about such behaviors, only then we can take corresponding actions not only to predict locations, but also to prescribe actions (e.g., offering more organic coffee) to improve visitor rates. 
This understanding of (human) behavior will broadly affect applications using mobility data. Traditional spatiotemporal data science allows for predictive analytics to predict the future. In contrast, mobility data science enables prescriptive analytics by understanding the underlying human behavior to devise actions and policies that aim to achieve desirable targets.

An open problem for understanding mobility behaviour data is the lack of labels or human annotation to provide insights on the actual observations. There are several other tricks that have been proposed, including cross-domain data fusion as well as developing interpretability mechanisms for machine learning or deep learning models. When geographical information is fused together with contextual features and social behaviours, not only location prediction can be improved, but also insights can be provided about the underlying visitor behavior~\cite{xue2021mobtcast}, even if no human-labelled data are provided about the mobility behaviors.

Therefore, explainability of AI and machine learning models that have underpinned many of such predictive behavior models remain an open challenge, especially since deep learning models are black boxes. One such approach for deep-learning-based models is disentangled representation learning, and a recent work~\cite{zhao2022measuring} shows that the disentanglement of latent spatiotemporal factors can assist the explainability of how the underlying latent factors learned by deep learning models are correlated. It can also be used for dimensionality reduction, and assist in few-shot learning cases.

\paragraph{Challenge 9. Visual Analytics}
Visualization and exploratory analysis of mobility data has long been a hot topic in visual analytics~\cite{and-VAMbook}. More recently, the trend turned to combining visualization with modeling and simulation to support decision making~\cite{Lee_congestion_2020}. This kind of research is by necessity application-oriented, while much less is done on developing more general ideas and approaches. 

One general research problem that has only been slightly touched in visual analytics but not systematically addressed is human involvement in real-time analysis of big mobility data. Is it possible to define realistic scenarios for involving human intelligence in big data analytics taking into account the cognitive limitations of human analysts with regard to the amount of information that can be perceived, speed of processing, and time required for analytical reasoning and contributing to the analysis process? Also how to combine computational methods of analysis, such as ML, with human expert knowledge and reasoning?
The involvement of human intelligence is limited to thoughtful data preparation, feature selection, parameter setting, and so on. It would be great to find ways to make more direct and effective use of human-possessed concepts and, particularly, knowledge of causal relationships. 
Hence, a grand research challenge for visual mobility analytics is to develop approaches to understanding and modeling mobility behaviors from low-level movement data, such as trajectories of moving entities.  

The following research problem is how to analyze behaviors after they have been extracted from elementary movement data and represented by appropriate data structures. A conceptual framework should be developed to enable defining the types of conceivable patterns of movement behavior. This will provide orientation for developing visualization techniques facilitating visual discovery of behavioral patterns, as well as algorithmic methods for detection of specified types of patterns. These techniques and methods should be incorporated in systems and workflows for analyzing the contexts in which various patterns take place and developing models for describing and predicting mobility behaviors depending on the context.

\section{Mobility Data Management Infrastructure}
\label{sec:management}

Classical data management systems have been designed for generic data types, where spatial and temporal data can be supported as new additional types. Yet, the core functionality of the data management engine does not acknowledge the spatial and temporal properties of mobility data. For example, mobility data calls for storing and querying locations of objects that evolve over time. The evolution can be in the location, the extent, and/or the properties of the object. The evolution can happen in discrete steps, e.g., check-ins, or in a continuous form. Thus, it is desired that the data management platform is able to represent the history, the current location, and possibly the near future of the moving object. Another example is classical index structures that are built with the assumption that the read workload is significantly higher than the write workload, and hence the index structure does not change often. Meanwhile, mobility data exhibits a different workload where the write workload (e.g., object location update) is significantly higher than the read workload, which makes all classical index structures simply not applicable to mobility data. A third example is that simple queries over mobility data, e.g., nearest neighbor search can be supported by classical data management systems by finding the distance between the user location and all other objects, sorting all objects based on that distance, and getting the closest one. This cumbersome approach is mainly due to the lack of having a specialized nearest-neighbor operator. Should we have one, that operator can seamlessly integrate with the query executor and optimize of a data management engine to efficiently support a pretty important query in most data mobility applications. Finally, a last example is that classical methods for scaling up data management in distributed environments rely on data distribution, mostly based on the data keys. This does not work well in scaling up mobility data as it is always desired to distribute mobility data in a way that spatially and temporally nearby objects are grouped together in the same cluster or computing node. This sections discusses current efforts and challenges of mobility data management.

\subsection{Efforts in Mobility Data Management}
\label{sec:statusquoDataManagement}

There has already been extensive research in all layers of mobility data management infrastructure. On the data modeling aspect, early models based on the constraint databases model aim to support simple moving objects (i.e., points), e.g.,~\cite{Grumbach98}. More complex data types (e.g., moving regions) have been supported by later models based on abstract data types, e.g.,~\cite{Guting2000} that is still being used in recent systems, e.g.,~\cite{zimanyi2020mobilitydb}. More recent efforts have been introduced to capture the semantics of trajectories of moving objects. Other models were also proposed to capture specialized modes of movement, including indoor environments, e.g.,~\cite{Jensen09}, network constrained, e.g.,~\cite{guting2006modeling}, fuzzy trajectories, e.g.,~\cite{Trajcevski04}, and detecting periodic moving patterns, e.g.,~\cite{Behr06, DBLP:conf/kdd/JindalGTLH13, DBLP:conf/edbt/ElfekyAE04, DBLP:conf/pkdd/BerberidisVAAE02}. In terms of indexing, tens of index structures have been proposed to support efficient indexing, storage, and retrieval for spatiotemporal data as either historical data, current locations, or continuously updated locations, e.g.,~\cite{DBLP:journals/debu/MokbelGA03, DBLP:journals/debu/Nguyen-DinhAM10, DBLP:journals/tsas/MahmoodAKBA18,MPA19}. This forms the infrastructure support for various spatiotemporal query processing techniques for various query operators over moving objects, including spatiotemporal range queries~\cite{MXA04}, spatiotemporal nearest-neighbor queries, e.g.,~\cite{DBLP:journals/pvldb/AlyAO12, DBLP:conf/gis/AlyAO15,  DBLP:conf/edbt/AlyAO15,xu2018,DBLP:journals/vldb/SilvaALPA13}, reverse nearest neighbor queries~\cite{benetis2006nearest}, skyline queries~\cite{HLO+06}, and scalable spatial and spatiotemporal joins, e.g.,~\cite{DBLP:conf/ssdbm/XiongMAHP04, whitman2019distributed}.
 
In terms of academic full-fledged systems, the SECONDO system has been introduced in the early 2000 as a comprehensive testbed for distributed moving object databases covering all aspects of data modeling, indexing, and querying~\cite{Gting2010SECONDOAP}. More recently, MobilityDB, implemented on PostGIS, has been introduced as a scalable system with a wider functionality on moving object databases~\cite{mobilitydb,zimanyi2020mobilitydb}. In terms of Big Data systems, ST-Hadoop~\cite{AMM18}, SUMMIT~\cite{AM20} and HadoopTrajectory~\cite{bakli2019hadooptrajectory} systems extend the Hadoop system to support spatial-temporal data, and trajectories, respectively, while other systems, e.g.,~\cite{DBLP:journals/pvldb/MahmoodAQRDMAHA15, DBLP:journals/tsas/DaghistaniAGM21,  DBLP:conf/gis/MahmoodDATBPA18}, extend the Twitter Storm distributed data streaming system to support streamed location data. 
TrajSpark~\cite{zhang2017trajspark}, Dita~\cite{shang2018dita},  and TrajMesa~\cite{li2020trajmesa}
extend the Spark system to support various index structures and query operations over trajectory data. SharkDB~\cite{wang2014sharkdb} extends in-memory column-oriented storage engines to support trajectories. In the open-source community and in industry, PostGIS~\cite{postgis} supports very basic trajectory functions, Oracle spatial supports streaming point data to capture real time mobility~\cite{OracleStreamIoT}, while Microsoft Azure \cite{10.1145/2996913.2996916} supports storing trajectory data in Azure table and utilizing Azure Redis for indexing. Distributed-MobilityDB \cite{bakli2020distributed} integrates in the one hand the trajectory data management of MobilityDB with a distributed PostgreSQL database to provide a distributed moving object database.

\subsection{Challenges in Mobility Data Management Infrastructure}

Though there is already a lot of work in various components of mobility data management infrastructure, there is an apparent lack of integrated systems that offer comprehensive functionality to end users, encapsulated in full-fledged systems that support mobility data science. Hence, the challenges in this section mainly focus on the system building aspect.

\paragraph{Challenge 10. Building Systems with Mobility Data in Mind.} Location data has almost always been supported in data systems as an afterthought problem. Many systems, e.g., Postgres, Storm, Spark, and Hadoop, have not been originally designed with location data support in mind. What typically happens is that spatial data types get augmented into tuple-oriented systems to support the location data type. For example, a restaurant tuple that describes various attributes of a restaurant is augmented with the latitude and longitude of the location attribute of the restaurant to support location services. Spatial indexes are provided to speedup the access to these attributes, and some accompanying spatial operators are provided to operate on the location attributes to provide location services, e.g., range or k-nearest-neighbor searches. While this approach works to some extent, systems coming out of such approach end up with sub par performance for spatial data, and hence for mobility data. Given the myriad of applications that rely on mobility data, it is important that systems are extended with native support for locations and mobility data. So, mobility data types and operations should be integrated in the core of these systems, and not to be considered as an afterthought problem. This can go through all kinds of systems, starting from database management systems that need to be spatially- and temporally-aware to support mobility data to scalable big data and NoSQL systems, where injecting spatial- and temporal-awareness into their core functionality will inherit their scalability to support scalable mobility data science.

\paragraph{Challenge 11. Location Data as First-class Citizens.} Having locations as core of mobility data calls for treating location data as a first-class citizen in a location data system that at the same time can be extended to support other data types~\cite{ARZ+22}. These location data systems can be presented as Location+X systems, e.g., as in~\cite{ARZ+22}, where the data types “X” can be keywords (e.g., to support spatial-keywords and tweets), graphs (e.g., to support road-network data), relational data (e.g., to support descriptions of spatial data objects), click streams (e.g., to support check-in data), document data (e.g., to support points of interest and documents that describe them), annotated trajectories (e.g., location + time + textual annotations), among others. In many location services, more than one data type X may need to be supported, e.g., a graph data type combined with a document or keyword data types, which calls for a multi-model-like data system. This gives rise to an eco-system where location is at the core with some form of an extensible multi-model data system that supports the multitude of data types “X”. However, current multi-model data system technology is lacking in several aspects. First, they do not support data streaming that is a cornerstone in mobility data due to the online streamed locations of moving objects. Second, we do not want to fall into the trap of adopting existing multi-model technologies that may affect location being a first-class citizen. However, the need for supporting multi-models in one seamlessly integrated location+X system remains a necessity. In addition to supporting location data via a native location+X engine, an ecosystem for mobility data would also include many important utilities to facilitate a broad spectrum of location service applications. From the input data side, to help navigate the vast amounts of available location datasets, and discover the right data sets for a given task, a location dataset lake infrastructure and location dataset discovery, cleaning, and integration facilities are needed. From the presentation side, a comprehensive visualization suite is envisioned to support visualizations for combinations of spatial and temporal data analytics on top of location data.

\paragraph{Challenge 12. Streaming, Batch, and Hybrid Workloads.} Motivated by the application needs, mobility data management need to support both batch and real-time data through all systems layers from digesting the data to analyzing and visualizing it. For example, a common requirement is to visualize the positions of a fleet of vehicles in real time, which only requires access to the most recent positions of the vehicles. Yet, at the same time, there is a need to perform batch analytics on the full trajectory of these vehicles (e.g., to assess whether the trajectories exhibit some unexpected behavior). Generally speaking, the need to have both real-time and historical data has led to the development of the data warehouse domain, where operational databases cover the real-time Online Transaction processing (OLTP) while data warehouses cover the historical Online Analytical Processing (OLAP). Since having two different systems for the two kinds of workloads is very costly, a new approach referred to as Hybrid Transactional and Analytical Processing (HTAP) has been recently proposed. However, mobility data exhibits pretty different workloads from other data, where streaming data is kind of dominant in terms of objects continuously streaming their new locations. Meanwhile, historical data is not of less importance and are continuously appended. While some efforts have been spend in the direction of write-optimized indexing for location data, e.g., as in~\cite{DBLP:conf/icde/Shin0A21}, 
more research efforts need to be spent to adopt the concepts behind HTAP systems to support the nature of mobility data.

\section{Mobility Data Privacy}\label{sec:privacy}


As we discussed in Challenge 1, mobility data privacy is a core problem in the mobility data science pipeline. Studies have shown that location data could reveal sensitive personal information such as home and workplace, religious and sexual inclinations~\cite{pyrgelis2017does}. 
As localization technology advances and extremely fine-grained location tracking is being enabled, it may even reveal products of interest in the stores we have visited, doctors we saw at a hospital, book shelves of interest in a library, artifacts observed in a museum, and
generally anything that might publicize our preferences,
beliefs and habits. 
Recent survey has shown that 78\% smartphone users
among 180 participants believe that Apps accessing their location pose privacy threats~\cite{chatzikokolakis2017methods}. 

While there are many privacy-preserving data collection and data analysis techniques developed for personal data, mobility data introduces unique challenges due to 1) spatiotemporal correlations in the mobility data which often results in increased privacy cost due to privacy composition for correlated data or downgraded utility for downstream applications, 2) complex location semantics (e.g., corresponding POIs of locations) and mobility behaviors (e.g., regular vs. one-time visit of a location) which existing privacy definitions may not be able to capture, and 3) diverse and emerging application scenarios such as contact tracing using mobility data for which existing privacy algorithms designed for aggregate data analytics are not suitable.  In this section, we briefly review existing privacy notions and techniques developed for location and mobility data and discuss several open challenges.

\subsection{Efforts in Mobility Data Privacy}
We categorize existing techniques in mobility data privacy into two main settings corresponding to our data pipeline: 1) local setting (data collection stage), and 2) central setting (data analysis stage).  In the local setting, the mobility service provider that collects mobility data is
assumed to be untrusted, hence each mobile user or entity can apply privacy-preserving mechanisms before the data is collected by the service provider.  In the central or global setting, the mobility service provider is assumed to be trusted and collects the raw mobility data.  The provider can apply privacy-preserving mechanisms for statistical analysis, and share aggregated data, machine learning models trained from the data, or synthetic data mimicing the original data to untrusted third parties.  

\noindent
\textbf{Local Setting.}  In recent
years, local differential privacy (LDP), the local variant of differential privacy, \cite{gu2020providing,cormode2018privacy} has become the de facto standard for
preserving privacy at data collection stage. Each user can perturb her
raw data using an LDP mechanism before uploading it to an untrusted server. Most existing mechanisms are designed to ensure utility for aggregate queries or analytics (e.g., frequency or density estimation) and requires the aggregation of the perturbed values from a large group of users, while the individual perturbed value may not provide much utility. Several works 
applied existing LDP schemes to location data but the utility is
poor \cite{kim2019workload, zhao2019ldpart}. Other works relaxed LDP to personalized LDP \cite{chen2016private}. Recent works developed improved LDP mechanisms for location data with better utility \cite{wang2022privlbs}. 

In addition to support aggregate data analytics, location based services (LBS) including range queries, spatial crowdsourcing, and the emerging contact tracing for pandemic control, require the precision of the perturbed locations themselves. Geo-indistinguishability (GeoInd) \cite{andres2013geo} relaxes LDP for location data which requires the locations to be indistinguishable only within a radius and the indistinguishabilty is scaled by their distances, providing better privacy utility tradeoff for LBS.  
Later works extended GeoInd to account for temporal correlations between consecutive locations of mobile users \cite{xiao2015protecting} and protection of customizable spatiotemporal activities instead of raw locations or trajectories \cite{cao2019protecting}. Other works applied the GeoInd mechanisms and variants for privacy-enhanced spatial crowdsourcing and contact tracing \cite{to2018privacy, da2021react}. 
Besides statistical privacy techniques, Private Information Retrieval (PIR) and secure multiparty computation (MPC) techniques have also been developed to allow LBS queries such as range queries and contact tracing without revealing individual locations \cite{ghinita2008private,al2019privacy, cho2020contact,reichert2020privacy}, but
are generally more computationally expensive and need to be designed for each different query.   

\noindent
\textbf{Global Setting.} Many works have applied differential privacy (DP) for computing and publishing aggregate mobility data. Compared to DP algorithms for tabular data, they typically exploit the hierarchical structure of locations and sequential patterns of trajectories to improve utility \cite{chen2006efficient, qardaji2013differentially,mir2013dp,acs2014case,shaham2022htf}. Some works also utilized the DP aggregates for task assignment in spatial crowdsourcing \cite{to2014framework}.  In practice, mobility data providers have started sharing aggregated mobility datasets with DP, esp. in response to the pandemic, such as Meta's population density maps and Movement Range maps, Google's COVID-19 Community Mobility Reports, and SafeGraph's Patterns \cite{crisisready2022}.
Other works have applied DP for training machine learning models using mobility data, for example, for location prediction \cite{AhujaGS20}. 
 Another line of work attempts to generate synthetic trajectories or mobility data based on raw trajectories with formal DP guarantees \cite{he2015dpt, wang2023privtrace}.  From the privacy attack side, recent works demonstrated the possibility of membership inference attacks on aggregate location data and linking attacks, and the defense power of DP against some of these attacks, reinforcing the need for ensuring rigorous privacy even for seemingly anonymous aggregate mobility data and machine learning models trained from mobility data \cite{pyrgelis2017knock,jin2019moving}.  

\clearpage
\subsection{Challenges in Mobility Data Privacy}
This section highlights open problems related to mobility data privacy, that needs consideration from the community.

\paragraph{Challenge 12. Threat Models and Privacy Definitions.}
The first challenge for mobility data privacy is the need to understand the threat models and adopt or define proper criteria by which to enforce privacy. We need to define first what needs to be protected (i.e., the sensitive information).  This may vary for different mobile users and applications. It may be the exact location coordinates of a user at a given time (most existing efforts focus on this). It may also be the association of a user with a sensitive place, co-location of two users (while it's okay for the users to reveal the exact location coordinates), or spatiotemporal activities of a user (e.g., stay at a place, or a trajectory). When defining privacy models and designing subsequent privacy mechanisms, there will (almost always) be attacks based on side channel information exploitation. While privacy notions like DP typically assumes the worst case which also means sacrificed utility, relaxed versions may be needed given specific threat models to enhance the privacy and utility tradeoff.     

Besides developing rigorous privacy enhancing mechanisms, it is equally important to understand the privacy risks and the empirical defense power of the privacy enhancing technology (PETs). While there have been some work on privacy attacks on aggregate mobility data \cite{pyrgelis2018knock}, more work is needed to understand what sensitive information may be revealed and reconstructed from mobility data based models, e.g., if membership inference attacks or feature reconstruction attacks \cite{shokri2017membership,fredrikson2015model} can be carried out, and potentially build benchmark attacks which can be used to audit the privacy risk of mobility data science systems and privacy mechanisms.

\paragraph{Challenge 13. Privacy and Utility Tradeoff and Other Factors.} When designing privacy mechanisms for mobility data collection and analysis, it is important to consider the utility of the privacy protected data for the downstream applications. For LBS (as typical in the local setting), the utility needs to be measured by the precision or accuracy of range queries for POI search, or contact detection for contact tracing (instead of how accurate the perturbed location is from the original location for which most algorithms following geoInd are focused on). Hybrid methods that combine DP and cryptographic techniques may be needed esp. for critical applications like contact tracing and public health \cite{cho2020contact}. For aggregate data analytics and machine learning applications using mobility data (in both local and global setting), the utility need to be measured by the accuracy of the statistics (e.g., frequency or density estimation for which most existing work focus on), the trained model, or the fidelity of the synthetic data. As a result, the algorithms need to be designed to optimize the corresponding utility and many remain an open challenge.  For example, existing methods for DP trajectory synthesization are mainly based on statistical models or low-order Markov models and perform well on some utility metrics \cite{he2015dpt, wang2023privtrace}.  While there are more powerful generative adversarial network (GAN) based models or diffusion models for generating more realistic synthetic trajectories \cite{LucaBLP23-Survey-Mobility, zhu2023difftraj}, ensuring formal DP for these models would result in deteriorated utility due to the complexity of the models. Designing methods for optimal privacy utility tradeoff remains an open challenge. 
  

In addition to the privacy and utility tradeoff, privacy enhancing technology may exacerbate bias in the data or learning algorithms.  Mobility data may have inherent bias as we discussed in Challenge 2. Data analysis algorithms may also have unfair performance for groups that are underrepresented in training data. It has been demonstrated that learning with DP could
exacerbate such unfairness, i.e. underrepresented groups suffer from worse privacy/utility trade-offs \cite{bagdasaryan2019differential}.  Research is needed to understand such impact on mobility data and design privacy algorithms to optimize privacy utility tradeoff while ensuring the fairness.

\paragraph{Challenge 14. Explainability and Societal Education}
Another important challenge of mobility data privacy is to improve the explainability of privacy definitions and mechanisms and communicate them to the stakeholders including mobile users (data contributors), mobility service providers, and data analysts. This is a general challenge for privacy enhancing technology, but more so for mobility data given the complex semantics of location information and diverse applications as we mentioned.  DP-compliant algorithms and location privacy models (such as Geo-Ind) as described earlier use privacy parameters to control the trade-off between  privacy guarantee and the utility of the private outputs. 
However, there is a significant gap between the theory and practice of DP: we lack principles and guidelines for choosing privacy parameters when collecting or processing mobility data using DP techniques in the real world. While the technology companies have employed DP in releasing the mobility datasets as we discussed earlier, the choice of the privacy parameter and the associated noise and uncertainty are often
not precisely specified or uniform across
companies. This makes it difficult for the downstream applications to quantify the uncertainty of the analysis result. 

The parameter $\epsilon$  of DP is mathematically defined but not well-aligned with the stakeholders' interests. %
Even for the same $ \epsilon $, the privacy guarantees could be different based on the different variants of DP and algorithms at hand. 
In addition, the $ \epsilon $ is not always linked to a specific privacy risk for the users (such as "the probability that an attacker can correctly infer my data") or a precise utility level for data analysts (such as "the accuracy of the DP-ML model"). %
To promote the adoption of mobility data privacy technology such as those based on DP, we should establish principles, design guidelines, and provide tools for explaining  DP's protection and limitation  from stakeholders' practical interests.
For example, we can help data contributors understand the privacy risk (such as membership inference attacks or reconstruction attacks) under different privacy parameters given a concrete DP algorithm;
we can also design efficient methods to visualize how data analyzers'  utility metrics (such as MSE or model accuracy) may change along with different privacy parameters for specific mobility applications.



\section{Mobility Data Science Applications}
\label{sec:applications}
Mobility data science used to be limited in the domain of transportation but recent technological inventions have created an abundance of mobility data, resulting in applications in many other domains of interest for society. Such applications leverage mobility data to understand, explain, and predict where moving entities such as humans, animals, or infectious diseases go, why they go where they go, and where they will go next. This section outlines broad applications of mobility data science to illustrate the recent landscape of mobility data science. 

\subsection{Traffic}
Traffic is a problem of global scale, as recognized by transportation science over a decade ago. Drivers in the United States spend 
6.9 billion of driving-hours
stuck in traffic and waste more than 11 billion liters of fuel per year according to INRIX~\cite{inrixscorecard}. Measured per-capita, people in Russia and Thailand spend 
even more time in traffic, while Brazil, South Africa, the UK, and Germany are only slightly behind the United States.
Leveraging mobility data science and understanding the underlying behavior of human participants concomitantly with different transportation modes, can enable more effective solutions to multiple problems at the heart of improving traffic management. Two main lines of research focus on: (1) traffic monitoring at an aggregate level, e.g., to help city administration, and (2) services that road users are getting. Existing work towards traffic monitoring include monitoring congestion \cite{li2007traffic}, assessing the safety of roads and intersections~\cite{maeda2016lightweight}, traffic prediction~\cite{LiYS018}, evacuation routing~\cite{zhang2015evacuation}, optimizing the public transportation schedules~\cite{richly2015optimizing}. 
Efforts on the services provided to road users include routing queries that balance the traffic across roads~\cite{de2016real}, help find drivers finding nearest facilities~\cite{KolahdouzanS04}, personalized routing \cite{LiGLG19}, eco-routing for minimizing greenhouse emissions~\cite{lin2014survey}, and enabling multi-modal trip planning~\cite{TKLG18}. 
But there are many open opportunities and challenges in using mobility data to improve traffic conditions. One example is devising accurate models for the dynamic scheduling of public transportation. Another example is the context-aware optimization of traffic signals -- e.g., incorporating the impact of additional flux of pedestrians in bus/train stations, to minimize the stop-and-go impacts for vehicles. A challenge of using mobility data science in the transportation domain is monitoring and reduction of emissions. Being able to quantify emissions (e.g., from transportation) is essential to accountability and reduction of emissions. 
Using data on emissions collected from in-situ sensors but also sensed remotely through earth observation (satellite) data will allow us to better understand the effects of e-mobility, better collective transportation, and infrastructure improvements.

\subsection{Urban Areas}
In 2018, 55\% of the world’s population (4.2 billion people) resided in urban areas, and this proportion is projected to increase to 68\% by 2050~\cite{un2018revision}. Urban areas are a focal point for mobility application as they introduce a variety of mobility modalities such as electrical vehicles~\cite{vazifeh2019optimizing} and bicycles and scooters with respective sharing programs~\cite{li2015traffic}. But by understanding how, where, and why people move in cities, outer suburban and regional areas, the demand for infrastructure and energy can be better understood~\cite{zheng2014urban}. Improving this understanding helps reduces urban inequalities in cities~\cite{nijman2020urban} such as access to high quality food~\cite{walker2010disparities} and healthcare~\cite{guagliardo2004spatial}.
Mobility data also helps improve urban safety by improving crime prediction~\cite{fu2018streetnet} and helping to recommend safe routes~\cite{shah2011crowdsafe}. 

A specific Urban mobility data science supports urban areas is through data-driven map construction~\cite{ahmed2015map} and updating of existing maps to account for blocked or new road segments~\cite{CLHYGG16} which is paramount in autonomous driving applications~\cite{macfarlane2016addressing}.

The real-time monitoring of urban mobility could result in \emph{situational awareness}, initially a term coined in defense applications, involving \textit{perception} of the environmental states using the surrounding data, \textit{comprehension} of the ingested data to understand the emerging situations, and \textit{projection} of future states and/or events that require predictive analytics. Mobility data provides critical components and insights into situational awareness in cities. When achieved, this applies not only to enabling robust critical infrastructures in cities but also to protecting them from harm, e.g., forest fires, earthquakes, and terrorist attacks. Many researchers 
use 
mobility data as 
input to enable situational awareness in cities as well as in airports~\cite{shao2019flight}.




\subsection{Health Informatics}
The spread of infectious diseases is a highly complex spatiotemporal process that is strongly tied to human mobility~\cite{hou2021intracounty} and human behavior~\cite{elarde2021change}. Many recent works have used human mobility data for data-driven epidemic forecasting as surveyed in~\cite{rodriguez2022data}.
A specific example of leveraging mobility data for public health is contact tracing, which refers to the process of tracking persons who may have come into spatial contact with an infected person, and subsequently collecting further information about these contacts~\cite{mokbel2020contact}. The feature-rich interaction, processing and localization/communication modalities of smartphone devices, have brought these to battle on the technological forefront and have curbed the fast spread of pandemics, like COVID-19. To this date, the community has proposed a wide range of contact tracing approaches, including opportunistic~\cite{RambhatlaZSSL22} and participatory approaches~\cite{DaA0S21} approaches as well as privacy-sensitive~\cite{ZaidiAS22}, decentralized~\cite{troncoso2020decentralized}, proximity-based (e.g., BLE, sound)~\cite{reichert2021survey}, and location-based approaches (e.g., Wi-Fi, GPS)~\cite{DaA0S21} approaches. However, a wide range of challenges remain unanswered, including methodologies to improve the penetration and adoption rates, alleviate privacy or expectation skepticism~\cite{bedogni2021modelling}, ubiquitous availability on low-end terminals as well as technological/psychological adoption barriers~\cite{bedogni2022location}, achieving cross-country interoperability with standard formations beyond recommendations, scalability/reliability and accuracy verification of engaged spatial technologies as well as lessons about effectiveness from real large-scale deployments.

Another specific health application for mobility data is elderly health monitoring. GPS-enabled smart-watch technology can be used to monitor the movement of elderly users~\cite{stavropoulos2020iot}. In particular, if the monitored user is showing early signs of dementia, her/his trajectories could show an abrupt change from her/his movement history~\cite{tolea2016trajectory}. For instance, a user who normally walks in a park then goes to a restaurant is found to only stay in the park for a substantial amount of time. Indoor sensors installed in the room can also be used to track whether an elderly person or a patient falls from the bed. Trajectory outlier analysis methods, together with gerontology knowledge, can be very useful for this kind of applications.

\subsection{Indoor Environments}
Indoor mobility data management has been described as a new frontier in data management~\cite{jensen2010indoor}.
But in addition to data management, large-scale indoor localization data also raises challenges in data collection, data analysis and data privacy.  Indoor data collection is an open research problem due to the non-existence of the indoor equivalent of GPS: a system that can provide the user location in any building worldwide. This is particularly important in applications related to emergency management and infectious disease contact tracing. Systems have been developed over the years to address this problem based on different data sources including WiFi signal strength and time of arrival~\cite{yang2015wifi}, cellular signal~\cite{rizk2018cellindeep}, Ultra-wideband~\cite{alarifi2016ultra}, ultrasonic~\cite{ijaz2013indoor}, magnetic tracking~\cite{shu2015magicol}, inertial sensors~\cite{harle2013survey}, among others. These novel data sources enable new applications in indoor navigation, contact tracing, indoor analytics, and evacuation management. 

Indoor data analytics allows to improve understanding of indoor behavior which has multiple benefits and applications, including for crowd management~\cite{ahmed2014finding}, retail and POI recommendation systems~\cite{ren2017analyzing}, and for optimizing energy use and improving sustainability in the long term~\cite{salimBAE}.
For example, by utilizing WiFi logs, Ren et al.~\cite{ren2018understanding} find strong correlations between behaviors and user demography (e.g., age, gender and visitor types), indicating that both indoor mobility behavior, in conjunction with online behavior, can be used to predict the underlying demography of the visitors. 

Occupancy behaviors are also highly linked with building management systems and controls~\cite{carlucci2020modeling}. By having a more accurate energy use estimation using indoor spatial and mobility data, in addition to historical energy consumption data, the performance of the buildings can be better optimized, towards achieving a more sustainable operations~\cite{dong2021occupant}. The responsible use of mobility behavior analytics, including indoor and outdoor mobility behaviors, strongly points towards the increased capacity for improving sustainable operations of buildings~\cite{salimBAE}, enabling net zero goals to be achieved.

\subsection{Marine Transportation}
According to UNCTAD, Over 80\% of the volume of international trade in goods is carried by sea, and the percentage is even higher for most developing countries \cite{UNCTAD}. Estimates say that the global shipping activity emits $3\%$  of the global emissions worldwide in 2022 \cite{iea_global_emission_report_2022}. These significant numbers, as well as the availability of large-scale ship trajectory data obtained from the automatic identification system (AIS) \cite{artikis2021guide} motivated a lot of research efforts on mobility data analysis for maritime. The stakeholders who seek benefit  of such analyses include the maritime authorities, environment officers, ship owners, port and canal managers, and the transport and logistic sector. 

One major challenge is to ensure the safety at sea, which splits down to the technical challenges of identifying positional anomalies\cite{riveiro2018maritime}, locating dark vessels (vessels that switch off their AIS devices)~\cite{MAZZARELLA2017110}, and cleaning location and identity spoofing~\cite{Afflisio21}. Additionally, an essential aspect is the detection of fishing activities to ensure the sustainable fishing practices \cite{chuaysi2020fishing}. Since vessels do not have fixed routes in the sea, research has also investigated the density of ship routes~\cite{xu2017}.

Multi-criteria routing using multiple optimization criteria including estimated time of arrival, fuel consumption, safety, and comfort has been increasingly recognised as an important path planning problem \cite{hinnenthal_robust_2010}. An optimization of ship routes could effectively lead to significant reductions of GHG emissions and contribute to the actions against anthropogenic global warming. The influence of ocean currents, waves, and wind on the course and speed of ships have been known for centuries. Used optimally, ocean currents lead to more efficient paths between two given ports. Ship route computation approaches that exploit the potentials of wind, wave and weather models aiming at minimize fuel consumption have been addressed by the marine science, maritime engineering and transportation community \cite{fang2015optimization}. 

Since green mobility is currently gaining huge attention, CO2 emission aware ship routing is expecting to get an enormous impact in economy, politics and society and provides very promising opportunities for the spatial and spatiotemporal database and mobility community. Marine transportation becomes particularly important in the scope of climate change (e.g., the advent of hydrogen/battery/fossil/atom hybrid vessels) as well as  digitization for new infrastructure-free localization technologies on-board. 

\subsection{Social Connections}
Location-based social networks (LBSNs) bridge the gap between the physical world and online social networking services~\cite{zheng2011location}. LBSN data capture both human mobility (in the form of check-ins to discrete points of interest) and a social network between individual humans. Combining mobility data and social networks, LBSN data finds many applications. A first application found in the literature was on modeling and describing human mobility patterns (e.g., \cite{cho2011friendship,noulas2011empirical}), analyzing these patterns (e.g.,\cite{cheng2011exploring}), and explaining why individual user choose locations and how social ties affect this choice (e.g., \cite{wang2014checkins}). Another application is that of location recommendation, which leverages check-ins of users and their ratings in the user-location network to recommend new locations to users~\cite{bao2015recommendations}. A closely related application area is that of location prediction (e.g., \cite{cheng2013you}), which predicts the future check-ins of users.
Another active research field in LBSNs analysis is friend recommendation or social link prediction (e.g., \cite{scellato2011exploiting}), which suggests new friends to users based on similar interests at similar locations, while also having similar social connections.
Other research topics concerning LBSNs include efficient query processing (e.g., \cite{armenatzoglou2013general}) finding user communities (e.g., \cite{yin2016discovering}), and estimating the social influence of users (e.g., \cite{wen2014exploring}).

This plethora of applications and research shows how mobility data in connection with social network data can be used to understand the social fabric that ties us together. A potential future application is using human mobility data to reinforce this social fabric by recommending social events and meetings to groups of people to help people find new friends, collaborators, sports mates, teachers, mentors, and family members.

\section{Conclusions}
This paper presented the current state of mobility data science pipeline in addressing the specific challenges of mobility data. A main question that this paper answered is how mobility data science is different from data science. The space and time dimensions in mobility data call for different methods of data acquisition, management, analysis, and privacy preservation which are not addressed by the common data science tools. Accordingly we surveyed the main problems that are currently being researched, we identified major research questions for the coming years, and described applications that lead to broader impacts of mobility data science. Co-authored by a diversity of academics and industry professionals, this paper also conferred a community effort to sketch the boundary of mobility data science as an interdisciplinary field and bring together a dedicated research community around the identified research challenges.

\section{Acknowledgements}
Mohamed F. Mokbel acknowledges the support of the National Science Foundation under Grants Number IIS-1907855 and IIS-2203553. 
Mahmoud Sakr acknowledges the support of the EU's Horizon Europe research and innovation program under grant agreements No. 101070279 (MobiSpaces) and No. 101093051 (EMERALDS). 
Li Xiong acknowledges the support of the National Science Foundation under CNS-2125530 and CNS-2041952. 
Andreas Z\"ufle and Taylor Anderson acknowledge the support of the National Science Foundation under Grant Number DEB-2109647. 
Walid G. Aref acknowledges the support of the National Science Foundation under Grant Number IIS-1910216.
Gennady and Natalia Andrienko acknowledge the support of the Federal Ministry of Education and Research of Germany and the state of North-Rhine Westphalia as part of the \emph{Lamarr Institute for Machine Learning and Artificial Intelligence} (Lamarr22B), and of EU in projects \emph{SoBigData++} and \emph{CrexData} (grant agreement no. 101092749).
Reynold Cheng acknowledges the support of the Hong Kong Jockey Club Charities Trust (Project 260920140), the University of Hong Kong (Project 109000579), and the HKU Outstanding Research Student Supervisor Award 2022-23.
Panos K. Chrysanthis acknowledges the support of the National Science Foundation under grant number SES-2017614 and of National Institute of Health under grant number R01HL159805.
Anita Graser acknowledges the support of the EU's Horizon Europe research and innovation program under grant agreements No. 101070279 (MobiSpaces) and No. 101093051 (EMERALDS).
Matthias Renz acknowledges the support of the German Research Foundation under Grant Number 290391021 and 491008639, the Helmholtz School for Marine Data Science (MarDATA) partially funded by the Helmholtz Association (grant HIDSS-
0005) and the Federal Ministry for Economic Affairs and Climate Action (BMWi) under Grant Number 68GX21002E.
Flora Salim acknowledges the support of the Australian Research Council (ARC) Centre of Excellence for Automated Decision-Making and Society (ADM+S) (CE200100005).
Maxime Schoemans acknowledges the support of the Fund for Scientific Research (FNRS) under Grant Number 40018132.
Yannis Theodoridis acknowledges the support of the EU's Horizon Europe research and innovation program under grant agreements No. 101070279 (MobiSpaces) and No. 101093051 (EMERALDS).
Song Wu acknowledges the support of the EU's Horizon 2020 research and innovation program under the Marie Skłodowska-Curie grant agreement No. 955895 (DEDS).
Jianqiu Xu acknowledges the support of the National Science Foundation under Grant Number U23A20296.

\bibliographystyle{abbrv}
\bibliography{literature_v2}

\end{document}